\documentclass[]{article}

\usepackage{graphicx}

\begin{document}

\title{Stochastic dynamics of a sheared granular medium}

\author{Alberto Petri$^1$ \and Andrea Baldassarri$^{2}$ \and Fergal Dalton$^{1}$ \and Giorgio Pontuale$^{1}$ \and Luciano Pietronero$^{1,2}$ \and  Stefano Zapperi$^{3}$}

\maketitle
\thanks{$^1$ CNR-Istituto dei Sistemi Complessi, Via del Fosso del Cavaliere, 00133 Roma, Italy \\$^2$ Dipartimento di Fisica, Sapienza, Universit\`a di Roma, P.le A. Moro 2, 00185 Roma, Italy\\$^3$ CNR-INFM, SMC, Dipartimento di Fisica, Sapienza Universit\`a di Roma, P.le A. Moro 2, 00185 Roma, Italy}

\begin{abstract}
	 We experimentally investigate the response of a sheared granular medium in a Couette geometry. The apparatus exhibits the expected stick-slip motion and we probe it in the very intermittent regime resulting from low driving. Statistical analysis of the dynamic fluctuations reveals notable regularities. We observe a possible stability property for the torque distribution, reminiscent of the stability of Gaussian independent variables. In this case, however, the variables are correlated and the distribution is skewed. Moreover, the whole dynamical intermittent regime can be described with a simple stochastic model, finding good quantitative agreement with the experimental data. Interestingly, a similar model has been previously introduced in the study of magnetic domain wall motion, a source of Barkhausen noise. Our study suggests interesting connections between different complex phenomena and reveals some unexpected features that remain to be explained.
	\end{abstract} 
\section{Introduction}
\label{intro}

The response of a granular material to shear stress
has been widely investigated in the past. The first relevant
report on the subject was perhaps that of Bagnold~\cite{bagnold},
pointing out a transition from sliding to stick-slip motion.
Subsequent work (see e.g.\ references in~\cite{dalton05}), 
was limited essentially to situations with more continuous or constant 
shear rate. However, we may observe that 
the response of a granular medium to an applied shear stress is often
non-continuous, but takes
place in the form of intermittent, erratic events, particularly when the
applied shear rate is below some threshold. Despite the fact that
this situation is frequently encountered in nature
(earthquakes are a large-scale example), systematic investigation
of the statistics of fluctuations seems limited.
One recent step in this direction was taken by Nasuno and coworkers \cite{nasuno} who analyzed the stick-slip behavior of a sheared granular medium, but considering a highly periodic motion regime.
Molecular dynamics simulations  of a similar setup had been previously performed by Thompson and Grest~\cite{thompson}.
The aperiodic regime was investigated by Dalton and Corcoran in a slightly different apparatus~\cite{dalton01,dalton02},
who presented the statistics of slip amplitudes, durations and energy dissipation,
hinting at the possible presence of scale free distributions.
In this paper we report on some results from our study of
the stick-slip regime of a sheared granular medium,
when deviations from peridic motion are strong.
We describe the experimental apparatus in Sec.~\ref{experiment},
from which we extract the statistics of the reaction
force (Sec.~\ref{PDF}), the distribution of which has some
notable features (Sec.~\ref{stability}).
Section 5 discusses the statistics of velocity
fluctuations and compares with the stochastic model
and Sec.~\ref{summary} contains a brief summary.

\section{The experiment}
\label{experiment}
\begin{figure}
\resizebox{0.75\columnwidth}{!}{%
\includegraphics{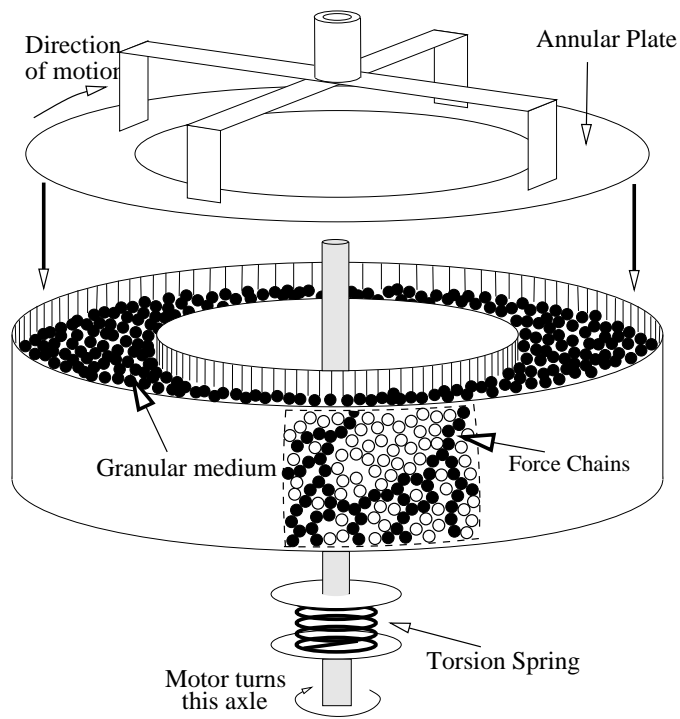}
}
\caption{The experiment consists of an annular top plate shearing over a
granular material in a Couette geometry.}
\label{fig:schematic}       
\end{figure}

In order to investigate the stick-slip response of a granular
medium to an external shear rate we devised and realized
a circular channel in which a granular medium is sheared
by an overhead rotating annulus which bears on the
medium (Fig.~\ref{fig:schematic}).
The plate is driven by a motor {\it via} a torsion spring,
and is free to move vertically.  As the motor turns
the springs loads and the annular plate is initiallly
retarded by static friction with the granulate.
As soon as the force is sufficient to destroy the granular
structure (i.e.\ force chains in the medium, see schematic in Fig.~\ref{fig:schematic}),
the plates starts to slip, unloading the spring,
and slows to a stop when the friction force overcomes the spring force.
Two optical encoders measure the angular difference $(\omega_D t-\phi)$ between the
motor and the plate, and this signal is stored in a pc as a function
of time. The resolution on the angle is $3\times 10^{-5}$ rad, and
up to $2\times 10^{-5}$s in time (though signals are typically smoothed and
interpolated to 10~ms or 1~mrad, in order to reduce instrumental noise).
The bed of glass beads is combed before each 
series of experiments
in order to minimize ageing effects and also to remove any
possible crystal structures present.
This preparation procedure 
allows for an acceptable reproducibility of the results.
More details on the experimental set-up can be found in~\cite{dalton05}.

\section{Statistics of the reaction torque}
\label{PDF}

\begin{figure}
\resizebox{0.75\columnwidth}{!}{%
\includegraphics{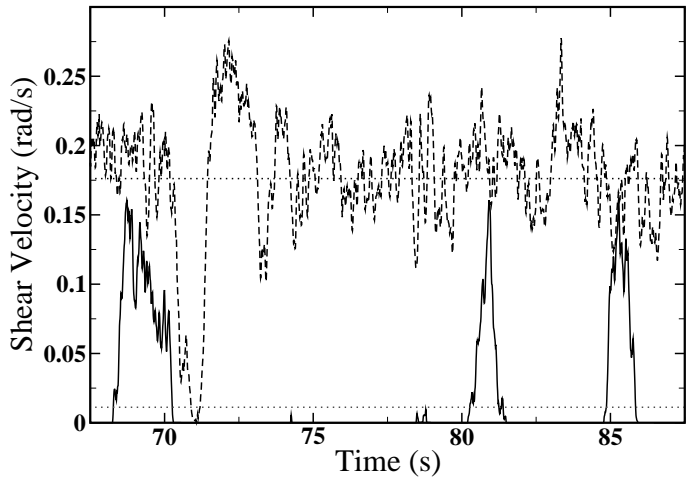}
}
\caption{Shear velocity as a function of time in the stick-slip (solid) and sliding (dotted) regimes. The respective driving velocities are shown by the dashed lines.}
\label{fig:shearvel}       
\end{figure}

By measuring the instantaneous angle, several quantities can be derived.
For instance, a sample of the resulting angular velocity time series
of the plate $\dot{\phi}(t)$ and driving motor $\omega_D$ is given in Fig.~\ref{fig:shearvel}.
The two curves with clearly distinct features correspond to the stick slip and to the sliding regimes.

The reaction torque $\tau$ of the granular medium can be derived as a functon of time:
\begin{equation}
\label{eq:motion}
\tau(t) = \kappa (\omega_D t-\phi(t)) + I\ddot{\phi}(t)
\end{equation}
where $\kappa$ is the spring constant and $I$ the total inertia of the plate and annexes.
The torque may also be more usefully expressed as a function of the angle: $\tau(\phi)$.
Experiments were performed at various driving velocities
and springs constants, the variation range of which is illustrated in Tab.~\ref{tab:1}.
The time series exemplified in Fig.~\ref{fig:shearvel} have been exploited to extract
important information on the statistical properties of the torque $\tau$.
In particular we have constructed the Probability Density
Function (PDF) $P(\tau)$ and the Auto-Correlation Function (ACF).
For each of the experiments, we have considered the following values of
$\tau$: $i)$ at the detachment (static), $ii)$ at stationary points of velocity 
(when $\ddot{\phi}=0$), and $iii)$ globally.  The resulting PDF's are shown in~\cite{dalton05}
and are virtually identical.  Their most notable feature, the asymmetry,
is clearly evident. Different functional forms can be used to reasonably fit
the curves (e.g.\ Gumbel, Lognormal, Gamma etc.) but none seem to offer any
insight into any possible underlying physical process~\cite{dalton05}.

\begin{figure}
\resizebox{0.75\columnwidth}{!}{%
\includegraphics{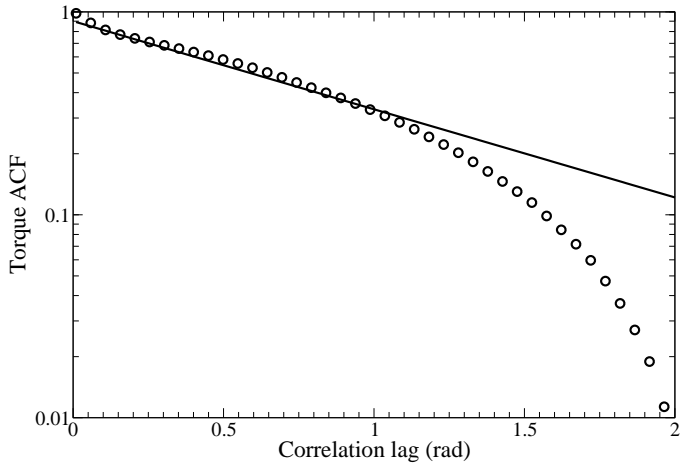}
}
\caption{The Autocorrelation Function for the torque exerted by the
torsion spring.  An initial exponential delay with characteristic angle
$\phi_c\simeq 1$~rad (solid line) is followed by a more rapid decay.}
\label{fig:trq-acf}
\end{figure}

The ACF is calculated on the torque signal as a function of
angular displacement, and in shown in Fig.~\ref{fig:trq-acf} for
a single experiment.  The straight line is an exponential decay
of angular constant $\phi_c = 1$ radian (corresponding to 30,000 times the
system's angular resolution, or roughly 100 particle diameters).
This observation suggests that a model for the granular friction force
should take explicitly into account the correlations present
in the granular medium.

\section{Stability and universality}
\label{stability}

Non-symmetric distributions appear in many natural phenomena,
and are usually deemed to be an indication of the presence of
correlations. In recent years some authors~\cite{unidist}
have suggested the idea that in many cases most, if not all, of
these distributions pertain to a simple universal kind of PDF,
which would take the place of the Gaussian distribution when
the variables are strongly correlated.
\label{sec:universal}

Actually, by virtue of the central limit theorem, the PDF for the sum
of a large number of independent variable converges to
the Gaussian distribution. In Ref.~\cite{unidist} and related work,
it is proposed that the observed non-symmetric distributions may be
limit distributions for sums of strongly correlated variables.
\begin{figure}
\resizebox{0.75\columnwidth}{!}{%
\includegraphics[width=0.9\columnwidth]{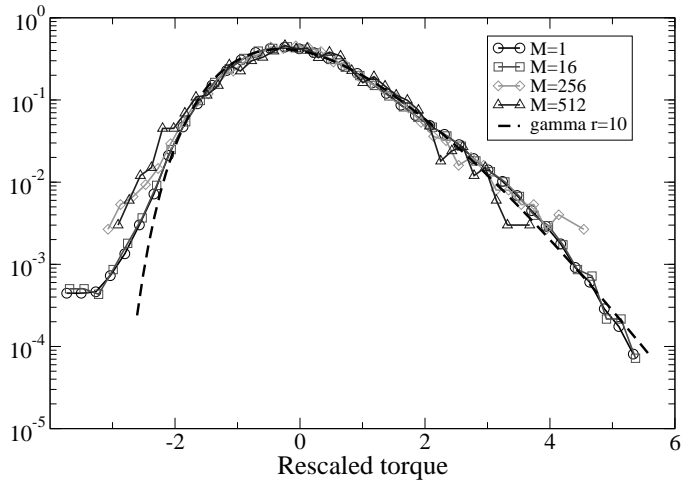}
}
\caption{Comparison of the sum variable $Z_M$'s PDF with rescaled gamma distribution with parameter $r=10$. The PDF of the torque corresponds to the  M=1 curve. It presents a pronounced asymmetry.}
\label{fig:gamma-comparison}       
\end{figure}

The distribution of the torque shows a very similar behavior, with a pronounced asymmetry. This is shown in Fig.4, where the torque distribution is the curve denoted by $M=1$ (see below).

In order to test if the torque PDF is compatible, at least in
principle, with such a limit distribution,
we have computed the PDF of the sum variable
\[
Z_M=\sum_{i=1}^{M} \tau(\phi_i)/M
\]
where $\phi_i= i \cdot \Delta \phi_0, \, \, (i=1,2,\dots)$, and we compare the distribution for
several values of M. 
In our data $\Delta \phi_0 \approx 1$ mrad. 
In figure  \ref{fig:gamma-comparison} we report the distributions for $M=1$,
$16$, $256$, and $512$ (i.e. corresponding to $1\rightarrow 512$ mrad).
The distributions are rescaled in order to
have zero mean and unit variance.

Recalling the analysis of the torque distribution, if we assume that
the variables $\tau_i=\tau(\phi_i)$ are Gamma distributed, i.e. according to:
\[
P(\tau_i) = \frac 1{\Gamma(r)\theta^r} \tau_i^{r-1}
\exp(-\tau_i/\theta)
\]
then for any $i$ the rescaled variable $z=\frac{\tau-<\tau>}{\sigma_\tau}$ should be distributed as
\[
P(z)=
\frac{\sqrt{r}}{\Gamma(r)}\exp(-\sqrt{r}z-r)(\sqrt{r}z + r)^{r-1}
\]
Fitting our measures for the torque distribution with a Gamma
distribution, we find that $r\approx 10$.
Note also that the variables $\tau_i$ are not independent: their
correlation coefficient decays exponentially as: $\rho^{|i-j|}$. A
comparison with the measured autocorrelation function gives
\[
\rho=\exp(-\Delta\phi_0/\phi_c) = \exp(-1/1000) \approx 0.999
\]

Accordingly with previous mathematical studies~\cite{kotz63,kotz64},
the distribution of $Z_M$ can itself be approximated with a gamma
distribution with parameter:
\begin{equation}
r(M) = \left( M^2  \left[ M + \frac{2\rho}{1-\rho}
\left(M-\frac{1-\rho^M}{1-\rho}\right)\right]^{-1}\right) r_1
\end{equation}
(the scale parameter is unnecessary if we compare the rescaled
variables). Considering our value of $\rho$, the variation of $r_M$
from $M=1$ to $512$ is  between $r_1=10$ and
$r_{512}\approx 11$. In Fig.~\ref{fig:gamma-comparison} we compare the
measured rescaled ditribution of $Z_M$ with the rescaled gamma
distribution ($r=10$).

Figure~\ref{fig:gamma-comparison} shows how the rescaled distributions do not change, and
this phenomenon can be interpreted as a form of ``weak stability'' of
the torque distribution \cite{renyi}.  A similar mechanism could also play a role in
other situations and may explain the universal behaviour discussed in Ref.~\cite{unidist,danube}.

\section{The equation of motion}
\label{motion}

\begin{figure}
\resizebox{0.75\columnwidth}{!}{%
\includegraphics{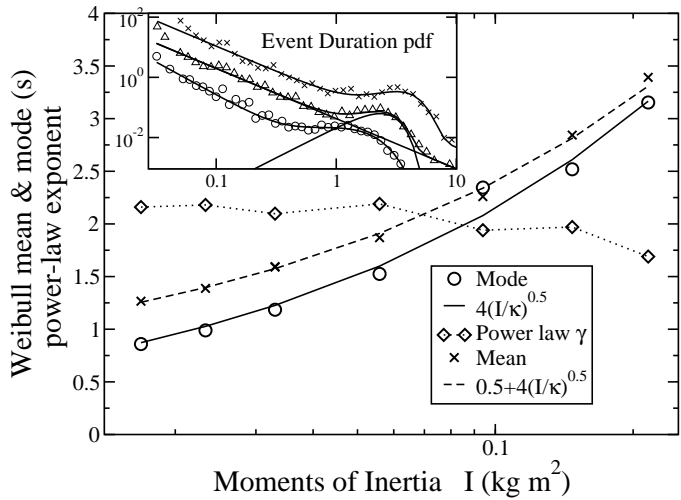}
}
\caption{The event duration distribution (inset) shows a power-law superimposed
with a peak, here fitted by a Weibull.  The mean and mode of the Weibull peak 
increase as the square-root of the moments of inertia, while the power-law exponent
remains constant.}
\label{fig:char-time}       
\end{figure}

The work by Dalton and Corcoran~\cite{dalton01,dalton02},
performed at constant volume,
revealed the presence of self similar PDFs for the
duration and extension of the slip events, and for
the velocity.
Our experiment confirmed the presence of such distributions, although in this case
the medium can dilate and a peak at large duration appears~\cite{baldassarri06} (inset to Fig.~\ref{fig:char-time});
this peak was ascribed to the inertial properties of the plate.
Figure~\ref{fig:char-time} shows the peak position as a function
of the moments of inertia of the top plate.  It is clear that the
peak displaces as $\sqrt{I}$ which is to be expected considering the top
plate as a harmonic oscillator with frequency $T=2\pi\sqrt{I/\kappa}$.
In the same article~\cite{baldassarri06}, we also presented
a stochastic model quantitatively describing the dynamics of the
stick-slip motion.

\begin{figure}
\resizebox{0.75\columnwidth}{!}{%
\includegraphics{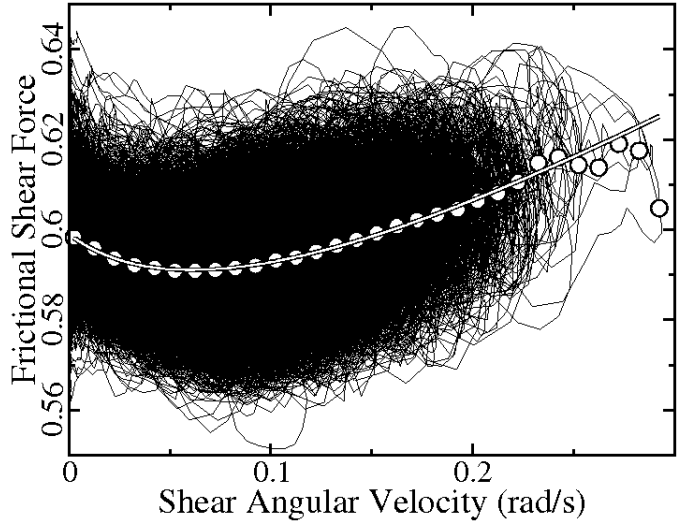}
}
\caption{The torque as a function of the instantaneous velocity for
each slip event, their average (circles) and the fit to Eq.\ref{friction} (thick solid line).}
\label{fig:trq-v-vel}       
\end{figure}

In order to find an effective description of the dynamics, an
extensive analysis of the reaction torque has been carried out. 
Figure \ref{fig:trq-v-vel} shows the reaction torque $\tau$ as a function
of the velocity for a set of slip events, apparently showing
a completely irregular behaviour. If an average is performed however,
then it is possible to identify a statistically regular behaviour.
Not surprisingly it is characterized by the usual dynamic instability
which is common to solid-on-solid friction and which is at the
origin of stick-slip motion. An adequate fit for the curve is
(Fig. \ref{fig:trq-v-vel}):
\begin{equation}
\label{friction}
\tau_d(\dot{\phi})=\tau_0 + \gamma (\dot{\phi}-2\omega_0 \ln(1+\dot{\phi}/\omega_0)).
\end{equation}
What is important is that it contains just two parameters, $\gamma$ the high-velocity
viscosity and $\omega_0$, the location of minimum friction, both of which
show little dependence on the experiment parameters $\omega$, 
$I$ and $\kappa$.

\begin{figure}
\resizebox{0.75\columnwidth}{!}{%
\includegraphics{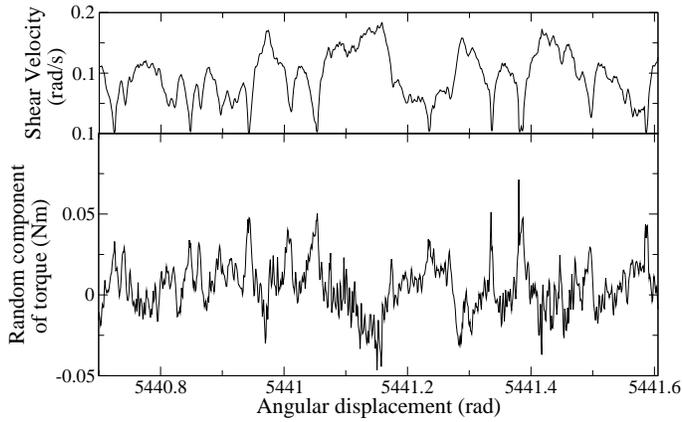}
}
\caption{The random component of the torque after the deterministic part, dependent
only on the instantaneous velocity, has been removed.  The upper curve shows the
velocity as a function of the angle. Events are delineated by moments when the velocity
reaches zero.}
\label{fig:trq-v-velb}
\end{figure}

Once  the deterministic part of the shear stress has been taken into account by the
friction law, depending only on the angular velocity, one remains with  only the stochastic 
part of the torque $\tau_s$ (which generally behaves as shown in Fig \ref{fig:trq-v-velb}).
Analysis of the correlation and of the spectral properties of $\tau_s$ suggests that,
with good approximation, it performs a bounded random walk in $\phi$ \cite{baldassarri06}:
\begin{equation}
\label{eq:randomwalk}
\frac{d\tau_s}{d\phi}= D\eta(\phi) - a\tau_s,
\end{equation}
$\eta$ being a unit white noise and $D$ and $a$ measured from the power spectrum
which should have the form 
\[
 S[\tau_s](k)=\frac{2D}{a^2+k^2}.
\]

With these few assumptions (Eqs. (\ref{friction}), (\ref{eq:motion}), (\ref{eq:randomwalk})) 
we are in a position to simulate the motion of the top plate.  The results have been outlined
in~\cite{baldassarri06} and the model shows excellent agreement with experimental
data such as the slip durations and sizes, and the velocity distributions.

Notably, it is quite interesting that an almost identical model
as that described here has been used to explain the origin of Barkhausen noise in
magnetic hysteresis cycles \cite{barkhausen}, where magnetic domain walls respond intermittently
to a varying external field.  Furthermore, through the use of this simple model,
chaotic granular stick-slip systems, heretofore notoriously intractable, have been opened
to analysis and characterisation.

\section{Summary and conclusions}
\label{summary}

\begin{center}
\begin{table}
\caption{Range of values over which the spring constant, $\kappa$, and motor angular
velocity, $\omega$ can be varied; in parenthesis values typically used in experiments.}
\label{tab:1}
\begin{center}
\begin{tabular}{ll}
\hline\noalign{\smallskip}
$\kappa (Nm/rad) $ & $\omega (rad/s)$  \\
\noalign{\smallskip}\hline\noalign{\smallskip}
0.12 (0.36) & $10^{-5}$ (0.005) \\
0.99 & 60 (1) \\
\noalign{\smallskip}\hline
\end{tabular}
\end{center}
\end{table}
 \end{center}

The stick slip regime of a granular medium subject to shear can be statistically
well characterized in terms of the probablity distribution of stress (torque),
velocities, slips durations and extensions, etc.

We measure the torque distribution, which is a physically relevant observable, and we observe
non Gaussian, asymmetric fluctuations. This distribution is similar to those encountered in
many different phenomena involving correlated variables and has been proposed as a sort of
'universal' distribution for sums of correlated variables.  We show that our results
seem to exhibit some degree of stability of the sum variable.  However, this result can be explained
by simply hypothesizing Gamma-distributed torque values, and so a true universal limit distribution is not implied,
particularly given the absence of a favourable convincing theoretical argument.

In our work we show that, despite the complexity of the mesoscopic interactions
between the grains, the statistics of fluctuations is quantitatively reproduced
by a minimal model, based essentially on the Brownian nature of the frictional granular forces,
accounting for the spatial correlation of the friction.

Both the experiment and the model exhibit broad slip duration distributions which
feature a characteristic peak.  In this work we present an analysis of this peak
and show, as expected, that its position is specified by the resonant frequency of 
the plate and spring.

This work was supported by the Italian FIRB project RBAZ01883Z and the EU project "TRIGS", 
c.n. 043386, COFIN 2005.
%
%
%

\end{document}